# Cyber-Physical Testbed for Power System Wide-Area Measurement-Based Control Using Open-Source Software

Hantao Cui [1], Fangxing Fran Li [1*], Kevin Tomsovic [1], Siqi Wang [1], Riyasat Azim [2], Yidan Lu [3], Haoyu Yuan [4]

[1] Department of Electrical Engineering and Computer Science, University of Tennessee, Knoxville, TN 37996, USA
[2] DNV GL, 2777 North Stemmons Freeway Suite 1520, Dallas, TX 75207, USA
[3] American Electric Power, 1201 Elm St, Dallas, TX 75270, USA
[4] Peak Reliability, 4850 Hahns Peak Dr., Suite 120, Loveland, CO 80538, USA
*fli6@utk.edu

**Abstract:** The electric power system is a cyber-physical system with power flow in the physical system and information flow in the cyber. Simulation is crucial to understanding the dynamics and control of electric power systems yet the underlying communication system has historically been ignored in these studies. This paper aims at meeting the increasing needs to simulate the operations of a real power system including the physical system, the energy management system, the communication system, and the emerging wide-area measurement-based controls. This paper proposes a cyber-physical testbed design and implementation for verifying and demonstrating wide-area control methods based on streaming telemetry and phasor measurement unit data. The proposed decoupled architecture is composed of a differential algebraic equation based physical system simulator, a software-defined network, a scripting language environment for prototyping an EMS system and a control system, all of which are integrated over industry-standard communication protocols. The proposed testbed is implemented using open-source software packages managed by a Python dispatcher. Finally, demonstrations are presented to show two wide-area measurement-based controls – system separation control and hierarchical voltage control, in the implemented testbed.

## 1. Introduction

Modern electric power systems rely on automated controls for secure and economic operation. The increasing level of renewable energy penetration through power electronics interfaces is bringing in faster dynamics and will require well-coordinated controls. Recently, synchrophasors have been deployed to obtain real-time data for situational awareness and fast, coordinated controls over a wide-area. Testing and demonstration of the effectiveness of wide-area measurement-based controls, therefore, become fundamental before actual implementations in real systems. Wide-area measurement-based control decisions are made based on global information which relies on the communication infrastructure. The conventional approach of modelling controllers using differential algebraic equations (DAE) has been dominant in power system simulation. However, this has many limitations in representing the measurement device characteristics and communication network behaviours. A more sophisticated simulation platform, namely, a testbed, needs to be designed and implemented to enable more testing capabilities.

A modern electric power system is intrinsically a cyber-physical system (CPS), which consists of the power loop as the physical part, and the communication loop as cyber. The physical part of a transmission system is an electric network of AC and DC transmission lines connected to generators, transformers, load, and power electronic devices. Electric power flows in the physical system following Kirchhoff's Law. The cyber part consists of a communication system in which information is exchanged, an energy management system (EMS, also known as supervisory control and data acquisition system, SCADA) in which generation scheduling is performed, and a control system in which fast wide-area control signals are generated. Information flows in the cyber system following defined information exchange routes.

The physical system and the cyber system are bridged by measurement devices and actuators. Measurement devices, such as synchrophasors, take time-stamped measurements from the physical network and send them to EMS/control systems for decision-making. Actuators, such as flexible AC transmission system (FACTS), receive the control signals and adjust the physical power flow accordingly. Therefore, closed-loop monitoring and control of a cyber-physical electric power system can be realized in an integrated fashion.

The major challenges in building a cyber-physical testbed for electric power systems are two-fold. On the one hand, in the physical system simulator/emulator, the mathematical equations to describe the power system dynamics need to be inclusive and have sufficient detail for control studies. On the other hand, in the cyber system simulator/emulator, the architecture, and topology of the communication network needs to be modelled to simulate delay, package loss, and cyber security events. Extensive efforts are required to design and implement a cyber-physical testbed from scratch.

Open-source software for the research community has captured growing attentions in recent years. Open-source software is distributed in the form of source code and can be reused or modified under license terms. For example, the most widely used operating system, Linux, and the most widely used web server, Apache, are both open-source software. In the power community, the number of open-source packages has also been increasing. Packages such as MATPOWER, PSAT, OpenDSS, GridLab-D, GridDyn and Andes are widely used for research activities. In the communication community, packages such as NS-3, OpenFlow, and Mininet are also open-source for communication network simulation. These packages are well maintained and documented.

The main contribution of this paper is the detailed design and implementation of a cyber-physical testbed, known as the Large-Scale Testbed (LTB), for wide-area measurement-based control verification and demonstration using open-source software. This paper is organized as follows. Section 2 provides a review on related work and available open-source software. Section 3 discusses the design choices and design architecture. Section 4 elaborates on the implementation of modules and the synchronization. Section 5 shows two sample case studies and demonstrations and section 6 provides conclusions.

## 2. Literature Review and Related Work

The interest in developing a cyber-physical simulation platform/testbed has been growing since the rapid development of the smart grid. Simulation has been a powerful method to study large engineered systems where it is difficult to perform actual tests; however, the traditional simulation method of the electric power system using standalone DAE are not sufficient for communication scenarios and cyber security assessments. The concept for designing the next-generation real-time control, communication, control and computation for large power systems has been proposed a decade ago [1]. Related works are categorized into the development of co-simulation platforms for power systems with communication, PMU-related cyber-security assessment, and wide-area measurement-based control applications.

The pioneer open-source software in the electric power society targets dynamics and economics simulation. Existing open-source software packages capable of simulating the physical system dynamics include PST[2], PSAT[3], MatDyn [4], and GridDyn [5]. They are positive-sequence phasor-domain transient simulators, which are essentially DAE solvers. The MATLAB-based simulators, PST and PSAT, have been widely used and are well tested by researchers. Specifically, PSAT has the most built-in power system model support in the open-source world to date. On the other hand, the recently unveiled simulator written in C++ language, GridDyn, utilizes a high-efficiency numerical solver, SUNDIALS, and has superior computational capabilities. The support for Modelica models in GridDyn is also under active development to interface open-source model libraries such as OpenIPSL [6].

Preliminary cyber-physical test platforms have been developed for smart buildings and smart grids for cyber security related studies, such as, intrusion detection, and network resilience. A toolkit for security research on CPS networks is proposed in [7] to connect CPS software and hardware, simulation scripts for components and physical-layer simulation engines based on network emulation, where case studies on cyber-attacks and defences are performed. A hybrid platform combining a distribution power system simulator and a software network emulator is proposed in [8] for testing impacts of communication network applications on power systems. A cyber-physical power system testbed is proposed in [9] for intrusion detection systems based on the real-time digital simulator (RTDS) platform and MATLAB. Architecture and studies of a cyber-physical security testbed, known as PowerCyber, is described in [10]. A global event-driven co-simulation framework is described in [11] for wide-area measurement and control schemes by showing a case study on communication-based backup distance relay protection scheme.

To represent the communication network with computer software, some of the works mentioned above utilize software-defined network (SDN) to set up and emulate reconfigurable communication infrastructures. The opportunities and challenges are assessed in [12] which discusses the benefits and risks SDN may bring to the resilience of smart grids against accidental failures and malicious attacks. SDN controller failures are studied in [13] to assess their impacts on the physical system by presenting an example of automatic gain control (AGC). In [14], a self-healing PMU network that exploits the reconfigurable feature of SDN is proposed to achieve resiliency against cyber-attacks, which is formulated into an integer linear programming model to minimize the overhead of the self-healing process. Cyber-attacks on energy-related CPS have also been studied in [15] for smart buildings, and in [16][17][18][19] for grid monitoring, protection, and control. Experience shows that open-source software-based network emulators are quick to setup and suitable for CPS testbeds aiming at fast prototyping and verifying power system research.

The above testbed designs can be taken advantage of for wide-area measurement-based control verification if the following challenges are addressed. First, telemetry and measurement devices, which describe the real-world counterparts need to be modelled. Second, the telemetry and measurement data should be streamed over industry-standard protocols, such as Distributed Network Protocol – 3 (DNP3) and IEC C37.118, to mimic the communication scenarios accurately. Finally, the measurement data may need to go through high-speed state estimation, which resides in the EMS system, for measurement-based control to reduce the impacts of measurement noise and errors.

This paper proposes a software architecture design for a cyber-physical testbed aiming at verifying and demonstrating wide-area measurement-based controls by leveraging open-source software. A decoupled architecture composed of power grid simulator, communication network emulator, EMS and control system is proposed. A dispatcher program calls components in the testbed as processes, which are connected to the SDN ports and exchange information using industry-standard protocols. Implementation of the testbed emphasizes open-source software for both the cyber and the physical part; however, the modularity allows commercial packages to be integrated if desired as well. Two measurement-based control case studies are shown to demonstrate the capability of the proposed cyber-physical testbed.

## 3. Design Architecture

The LTB is designed to mimic an electric power system with communication and control functions. The requirements for the design architecture are interchangeability and flexibility. Interchangeability means that the routines and functions are interchangeable with other routines and functions that have the same functionality. Flexibility ensures that the testbed is not limited to specific software packages originally supported but also connectable to other software platforms.

The design philosophy is to build a decoupled testbed and 'glue' modules together through data streaming. Each function or routine can be encapsulated as a module, exchange data, but also remain ignorant of other module structures. Interchangeability is realized when a module, for example, a power grid simulator, is replaced with another simulator producing the type of output. Flexibility is also realized when the testbed supports standard data streaming protocols which most software platforms can easily support.

The architecture of LTB as a cyber-physical testbed is depicted in Fig. 1. LTB is composed of four types of components: a power grid simulator to calculate the physical power network, a SDN to emulate the communication network, conventional EMS/SCADA with functions to control generation and load and an operator interface, and wide-area measurement-based controls to perform fast automatic controls

### 3.1. Decoupled Framework

The decoupled architecture is a framework that integrates the cyber and physical components into one testbed while allowing independent development of components. In the LTB, each component is a software module that performs a specified task given the defined input and output, such as running a power grid simulation, handling network connection and data transfer, or running wide-area measurement-based control. One module does not need to know the details of other modules if their communication interfaces are unified, for example, using the same communication protocol.

The decoupled modules in the framework are integrated using industry-standard communication protocols over a TCP/IP network. The communication protocols from the power industry applied are DNP3 for SCADA controls and IEC C37.118 for PMU data streaming. From the power grid viewpoint, measurement data is collected from the power grid simulator for processing in the EMS and control systems. From the communication perspective, DNP3 protocol in the application layer utilized for transferring telemetry data and sending control signals over TCP/IP in the software-defined communication network. On the other hand, IEC C37.118 protocol is used to stream PMU data from the power grid simulator to Phasor Data Concentrators (PDCs).

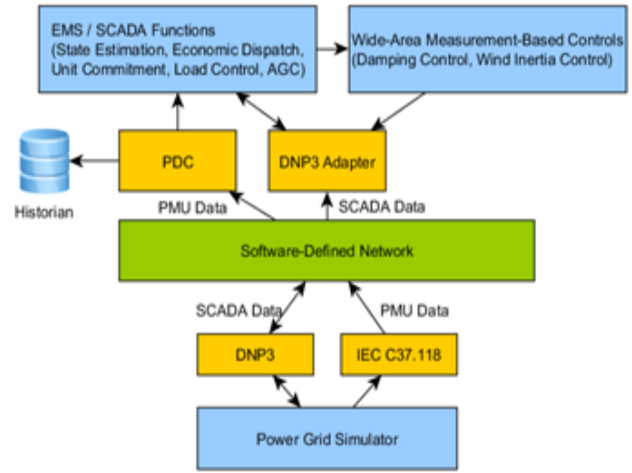

Fig. 1. Design architecture of LTB incorporating power network, communication network, EMS and control

### 3.2. Power Grid Simulator

The power grid simulator runs computational routines to simulate the characteristics of the physical electric power grid. The main component in the physical network include generators, transformers and transmission lines, shunt reactors and capacitors, various distribution system components, and loads, which together can be hundreds of device types. The time scale of interest for dynamic power system simulations can vary from microseconds, for say lightning or switching problems, to several hours, for say scheduling or restoration problems

There has always been a trade-off between modelling detail and simulation time, which arises in a transmission network simulator with potentially tens of thousands of buses. For transient stability analysis ($10^{-3} \sim 10^1$ s) of large-scale systems, positive-sequence phasor-domain models are the de facto standard in both commercial and open-source software. This assumes a balanced network, ignores transmission line transients and considers only the positive symmetric component. This greatly reduces the computation burden and is sufficient for electromechanical transient studies.

The power grid simulator produces output at each integration step from integrating the DAE. In a real power grid, data are acquired by telemetry and measurement devices, which only have a certain level of accuracy. To describe measurement errors and noise, a software module is added to modify the precise data from the simulation and impose errors and noise based on studies of the measurement model [20]. Such measurement data are then sampled for transmission using DNP3 and IEC C37.118 protocols at their specified sampling rate, respectively, before sending over the communication network.

### 3.3. Software-Defined Network Emulator

The SDN emulates the communication network on which the data packages are transferred. The SDN is composed of physical links, switches, and routers, and is configurable in software to emulate different network scenarios. In the decoupled framework, each module with communication capability is a host on the network and has virtual ports. Each host can run processes in its namespace, and send data over the emulation network. Hosts in the SDN include the telemetry sensors, PMU, EMS / SCADA modules, and the measurement-based control modules.

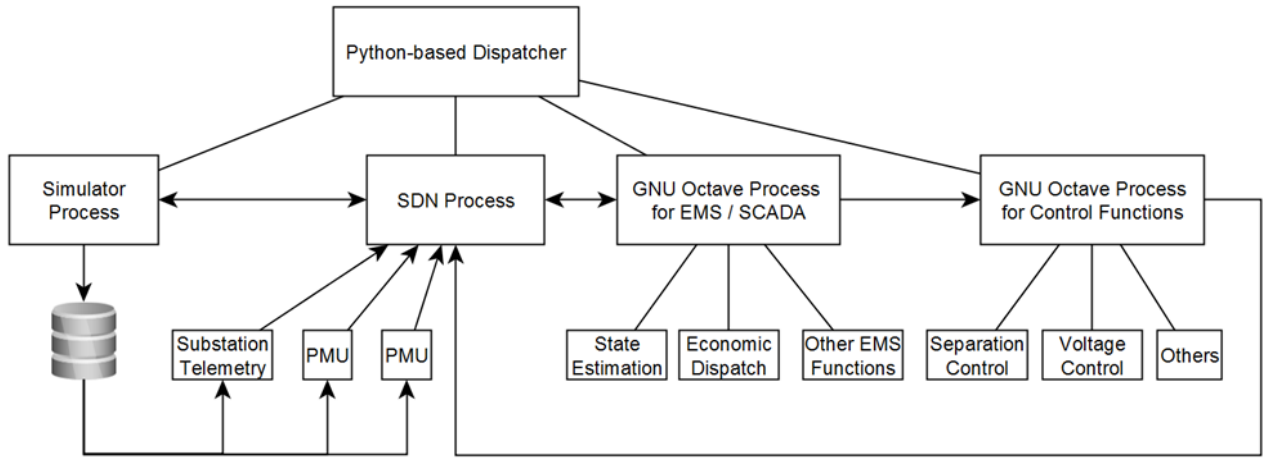

Fig. 2. Process dispatch architecture in the proposed implementation

The advantage of integrating SDN into the cyber-physical testbed is simplicity. No source code modification is needed to migrate from a real network port to a virtual one. Each host can link to a real port or a virtual port simply by changing the port name and writing to sockets as usual. Network latency and package loss scenarios can also be represented by modifying the software definition of the network. Some SDN allow binding physical ports in the virtual network so that cyber security scenarios can be studied. Drawbacks of using SDN in a testbed exist and have been discussed in literature. The first concern is performance where thousands of virtual hosts, such as, for a continent-wide communication for a power system, may require large amount of computing resources. As a consequence, the data flow may suffer longer latencies in large virtual networks due to resource limitations.

### 3.4. EMS/SCADA Functions

The EMS / SCADA system plays an important role in power system energy balance scheduling and minute-level control. An EMS system gathers telemetry data, forecasts load, runs unit commitment and economic dispatch, screens contingencies, state estimation, and sends automatic generation control (AGC) signals. It is the operator interface for running the power system. The time scale of scheduling and control in an EMS system varies from day-level (unit commitment) to hour-level (economic dispatch) and second-level (AGC). Most existing EMS / SCADA systems communicate using the DNP3 protocol. A modification is to integrate high sampling speed PMU data so that near real-time data can be utilized for situational awareness. A high-speed state estimation module in the EMS is also needed to process PMU data, which serves as the foundation for wide-area measurement-based control functions. Other conventional controls in the EMS, such as economic dispatch and AGC, can also be updated to adopt PMU data. In the LTB, the EMS / SCADA system may have multiple implementations running simultaneously in the decoupled architecture by adding a data splitter after the DNP3 converter and the PDC. This is helpful to evaluate and compare results of different methods applied to the same system.

### 3.5. Control Functions

The control functions block is an aggregation of wide-area measurement-based control methods developed by researchers. They accept measurement data and state estimation output from the EMS /SCADA system, monitor critical variables and generate control signals. Control signals are sent back to the power grid simulator using the DNP3 protocol over the communication network. These measurement-based control function can be either centralized or distributed, depending on the information they request from the EMS / SCADA system.

The time scale of the measurement-based control functions is from sub-second to the second level as they are designed to handle faster dynamics in the system than the EMS / SCADA functions. For example, power electronic devices in a FACTS device may require fast control with inter-area information, and thus, can be included in the wide-area measurement-based control block.

### 4. Open-Source Implementation

The implementation of the proposed cyber-physical testbed, LTB, using open-source software is described in this section. The implementation comprises of the power grid simulators, SDN emulator (Mininet), telemetry and measurement units (OpenDNP3, pyPMU), EMS / SCADA functions (GNU Octave), and measurement-based control functions (GNU Octave).

### 4.1. Decoupled Design

To realize the decoupled design, a dispatcher program is created for managing the module processes and handling their synchronizations. The dispatcher is written in Python as it is an extremely flexible open-source scripting language, which has gained huge popularity in recent years. Python can interface to code and programs written in other languages and provide application program interfaces (APIs) through their Python bindings. Moreover, Python has a simple and elegant syntax and a flat learning curve. Therefore, Python is ideal for a dispatcher program that handles process creations and synchronization.

The Python-based LTB dispatcher creates processes for a) each telemetry device with DNP3 client support, b) each PMU device with IEC C37.118 support, c) an EMS /SCADA system with DNP3 master support, and d) a wide-area control system with ZeroMQ support. More specifically, each telemetry device loads data from the simulator, imposes measurement noise and error, and sends data to the DNP3 server, at a sampling rate of 1 frame every 2 seconds though the socket bind to a virtual port. Each PMU device loads data from the simulator imposes noises and errors, and sends data to the PDC, at a sampling rate of 30 frames per second, through the socket bind to a virtual port. The EMS / SCADA process runs a DNP3 server and a PDC, maintains a list of functions that are called routinely in GNU Octave to process the incoming data. The measurement-based control process listens on a ZeroMQ socket to receive the measurement data and state estimation results and calls control routines in GNU Octave to generate control signals.

Software packages used in the dispatcher program are a) OpenDNP3 for telemetry devices, b) pyPMU for PMU device and the PDC program, c) GNU Octave, which is a free and open-source MATLAB alternative, for EMS / SCADA and control functions, and d) ZeroMQ for data streaming between the EMS / SCADA process and the control process. Fig. 2 shows the implemented process dispatch diagram, where the arrows stands for the direction of information flow, while the solid lines without arrows represent the process hierarchy.

### 4.2. Power Grid Simulator

Although positive-sequence phasor-domain simulation for transient stability simulation in power system has been standard for many years, the choice of an open-source power system transient simulator is relatively limited. Reference [21] records a list of existing open-source power system simulators, in which only a few support DAE-based transient simulation. These include Power System Toolbox (PST), Power System Analysis Toolbox (PSAT), MatDyn, GridDyn, and our in-house tool, Andes.

The choice of power grid simulator depends on the model library and flexibility of the software. The model library determines the initial capability of modelling power system devices, and flexibility determines the convenience of adding new device models. In LTB, PSAT, GridDyn and Andes are interfaced and made available as the simulators, all of which are open-source. Each of them has distinctive features: PSAT has native GNU Octave support, and has a considerable number of built-in models; GridDyn is written in system programming language and utilizes the SUNDIALS [22] package as the underlying solver, and is, thus, computationally efficient. Andes is an alpha-release in-house package written in Python with an aim to provide rapid model prototyping, advanced model interfacing, and data analysis capabilities.

Owing to the decoupled design, the user can choose the power grid simulator based on the actual needs, such as, model support, simulation speed or data analytic interfaces.

### 4.3. Software-Defined Network Emulator

The SDN process in the LTB is based on Mininet [23] to create a software-based network for emulating a real communication network with switches. Mininet is installed on a Linux system, and Python APIs are used to create virtual network configurations and virtual ports/interfaces. Each interface is connected to a component, such as, a PMU or an EMS / SCADA system, described in Section 4.1 to mimic a real power system where the measurements and control system are distributed and connected to the network. The SDN process is responsible for creating virtual network topology scenarios, and virtual Ethernet ports in Linux, *veth*, which are linked to the virtual network. After initialization, the SDN emulator is ready to serve network connections and operate identical to a real network.

### 4.4. EMS and SCADA Functions

The EMS / SCADA process is composed of a data gateway and a set of data processing functions. The data gateway process runs a DNP3 server to gather data from telemetry devices, and a PDC server to gather data from PMUs. The process also keeps a list of EMS / SCADA functions to be called each time new data arrives or at a given time interval. To utilize the existing EMS / SCADA functions written in MATLAB, the EMS / SCADA process in LTB leverages Oct2Py package which calls GNU Octave from inside Python. Understandably, this approach could be improved by rewriting the MATLAB code in Python or using Python port of the desired packages. For example, for optimal power flow based economic dispatch function, a Python port of MATPOWER, PyPower, is interfaced instead of calling MATPOWER in GNU Octave. The port to Python is, however, not mandatory since MATLAB functions executed in Oct2Py are fairly efficient.

Implemented EMS / SCADA functions to date include: a) economic dispatch based on PyPower, b) unit commitment written in GNU Octave, c) contingency screening based on PyPower, and d) state estimation written in GNU Octave. In the decoupled architecture, each component function in the EMS / SCADA can be replaced with another program that has the same functionality. For example, the traditional weighted least-square state estimator originally implemented has been replaced by a two-stage dynamic state estimator [24] which estimates not only the voltage phasors but also the relative generator rotor angles.

Scheduling and control outputs are sent to the DNP3 adapter and transmitted to the power grid simulator. Meanwhile, the state estimation output and the measurement data are stored in Python objects and transmitted to the control function data gateway, using ZeroMQ through TCP/IP, as inputs to the wide-area measurement-based control modules.

### 4.5. Control Functions

Similar to the EMS / SCADA process, the control process in the testbed is a collection of experimental control functions that take in wide-area measurement data and compute control actions. Given the fact that most experimental control prototyping is done in MATLAB, the control process interfaces to GNU Octave through Oct2Py to use the research code. The control process is also composed of a ZeroMQ data gateway, which receives measurement data and state estimation data from the EMS / SCADA system, and a list of control functions to call upon data refreshment. Note that the measurement data comes in at a higher rate than the DNP3 telemetry data. Therefore, the control functions are also called more frequently, which will be more efficient if the control functions can be rewritten in Python.

Local controls and wide-area controls are implemented based on the same approach, where the only difference lies in the measurement data visibility. In the DAE-based power grid simulator, there is no strict boundary between a utility company owned local network and an ISO-operated large power grid. Therefore, the control functions are categorized into local controls and wide-area controls depending on the input data it requests from the EMS / SCADA system. This approach shows simplicity by sharing the same data channel but using different subsets to implement both local and wide-area controls and allows testing of various control system architectures.

The implemented wide-area measurement-based controls in the LTB to date include: a) under-frequency load shedding control, which monitors the system frequency and reduces load in severe low-frequency conditions, b) system separation control, which monitors severe line faults and estimated generator rotor angles and separate the system into islands as a remedial approach [25], c) hierarchical voltage control, which controls the bus voltage from three time scales: wide-area generation dispatch, regional var regulation, and local excitation, and d) online voltage stability assessment using a tangential index from Thevenin equivalent [26]. The control signals are sent to the power grid simulator using DNP3 protocol through the SDN

**5. Case Studies**

The developed cyber-physical testbed is used to verify and demonstrate two wide-area measurement-based controls: system separation control for emergencies and hierarchical voltage control. Test cases are based on a reduced WECC system with wind scenarios developed in CURENT [27]. The WECC system contains 181 buses, 313 lines (including 48 transformers), and 31 generators in the summer peak load scenario. The phasor measurement data on all the buses are aggregated into one PMU using multi-stream and sent across the emulated point-to-point network.

The results are visualized in a web-based visualization tool, which is developed mainly for research and demonstration. The visualization tool is composed of a Python-based data ingestion module which reads data from the simulator through ZeroMQ sockets, and a database in the back-end. In the front-end, it consists of a JavaScript-based cross-platform web application using Leaflet library and OpenStreetMap to draw buses, lines, and other physical system components, and render the contour triangulations for the selected variables.

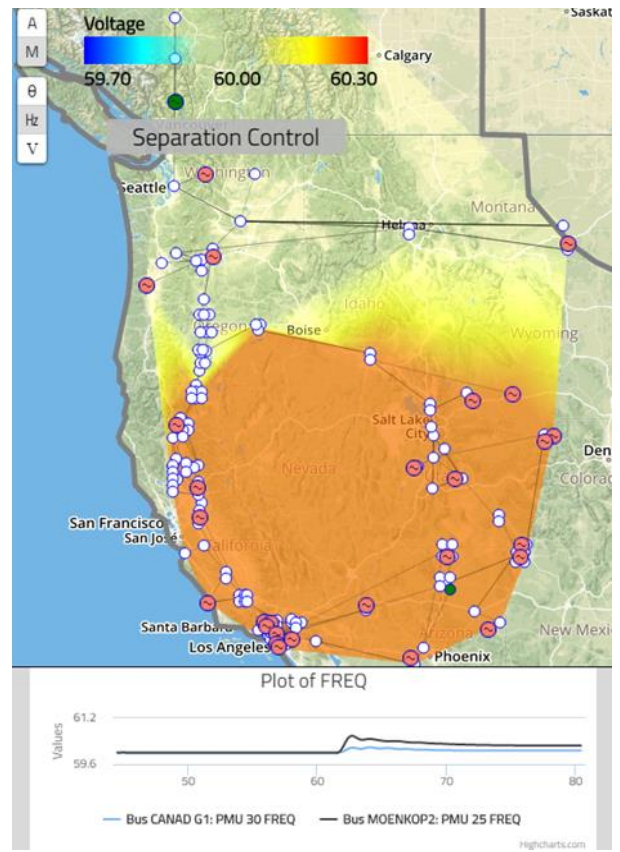

Fig. 3. Visualization of WECC 181-bus system frequency contour map after system separation

One capability envisioned for the visualization tool is to compare cyber-physical simulation scenarios simultaneously. To enable the simultaneous comparison, the visualization tool employs an HDF5-based data storage to store simulation data and a PostgreSQL-based database to store metadata. Side-by-side simultaneous comparison of scenarios are implemented by loading the desired scenarios from disk, cache them in the memory with *memcached*, and fetched from a web socket server.

*5.1. System Separation Control*

The system separation control splits the system into multiple islands during an emergency. When multiple lines are disconnected, the connection between two or more areas of a system becomes weak. Under such conditions, angular instability issue may arise. The objective of controlled separation scheme is to separate the system into several islands, and therefore, maintain the stability of each island. It is a wide-area, interconnection-level control method that utilizes generator rotor angles estimated from the two-stage dynamic state estimator in the EMS / SCADA system.

The location of separation is based on offline studies on the test system, in our case, the WECC system. Several elementary potential islands (EPI) and separation locations are studied and defined for WECC system in [25]. The first step for implementing this control scheme is to decide the best locations to divide the system. Based on the study of WECC's

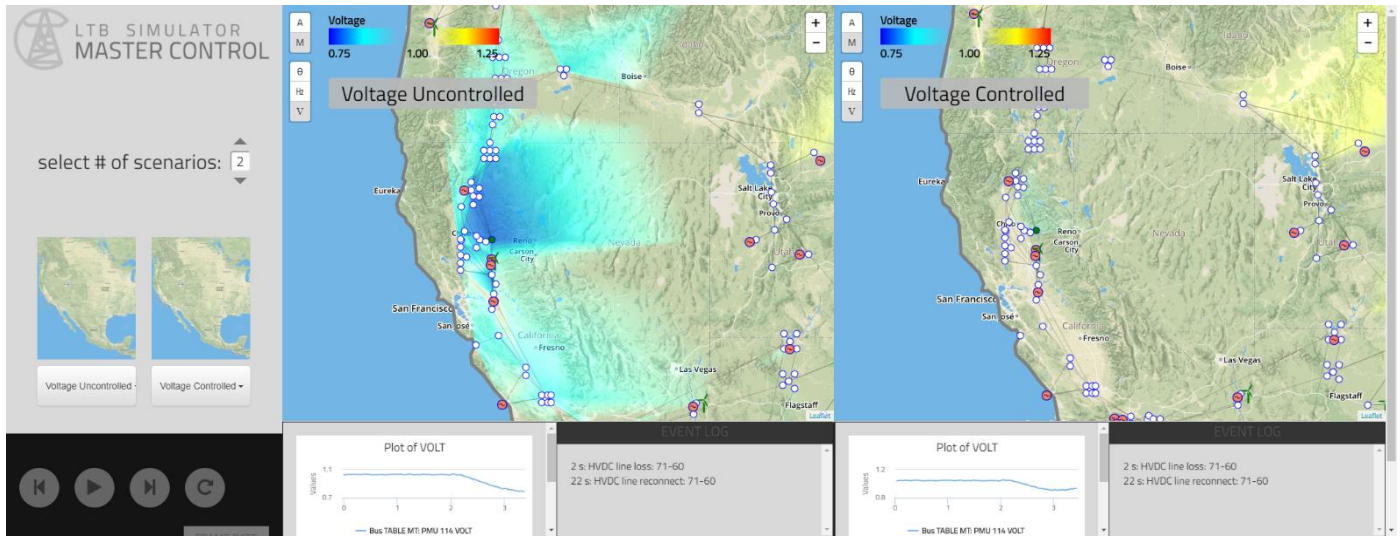

Fig. 4 *Simultaneous comparative visualization of hierarchical voltage control on WECC 197-bus system*

elementary coherent groups, 4 subsystems are defined as the EPIs, which can also be found in [25].

The timing of separation is also an important factor to consider. At each time step, the difference between the mean values of the largest 10 rotor angles and the smallest 10 rotor angles is calculated, when the absolute value of this difference reaches a certain threshold, the system will separate into islands. The control module communicates with the grid simulator through a ZeroMQ socket.

Lines 83-172 and 83-170 are disconnected at 5s and 40s, respectively, to create a weak connection between EPI 0 and EPI 1. At 61.5s, the system is separated into two subsystems by cutting the interface EPI 0 / EPI 1 and the interface EPI 0 / EPI 3. The load is reduced 10% in EPI 1 to maintain stability of the island comprised of EPI 1, EPI 2 and EPI 3. The frequency contour map from the simulation result is visualized in Fig. 3, which demonstrates the successful system separation.

### 5.2. Hierarchical Voltage Control

The hierarchical voltage control allows wind farms to support the interconnection-level system-wide voltage control and reactive power dispatch. The control aims at maintaining reactive power sufficiency in systems with high level of wind power by hierarchical measures: a) re-dispatching power flow in the interconnection level, b) regulating regional wind farm reactive power, and c) tuning local excitation system parameters for generators. The interconnection-level re-dispatch is implemented as an economic dispatch routine maximizing reactive power support, while the other two controls are realized in the control system. This control combines local control and wide-area control and utilizes voltage phasor estimation from the EMS / SCADA system.

The hierarchical voltage control is tested on a modified WECC 197-bus system with 22% wind penetration. The contingency to demonstrate the proposed control is a monopole DC loss, which heavily stresses the California-Oregon Intertie. NERC standard recommends that any voltage dip lower than 0.8 pu for more than 40 cycles is considered a voltage instability, which may trigger the actions of Under Voltage Load Shedding (UVLS). The simultaneous comparison of the controlled and uncontrolled scenarios is shown in Fig. 4.

The voltage contour map clear shows the low-voltage region in the system due to the HVDC line loss. The implemented voltage control method, which is based on measurement data, is effective in reducing voltage dips in the studied system.

### 6. Conclusions

This paper proposes a cyber-physical testbed design and implementation for wide-area measurement-based control using open-source software. A positive-sequence phasor-domain power grid simulator is adopted for the physical system, and a SDN emulator is used to establish the cyber system. A decoupled architecture is proposed by connecting telemetry and measurement devices over the communication network to an EMS / SCADA system and a control system over standard data streaming protocols. In terms of implementation, the functional modules are dispatched by a Python-based program on a Linux system.

The LTB platform based on open-source tools is designed for testing wide-area measurement-based control techniques. It allows for quickly integrating modern EMS functions and measurement-based control methods in large-scale cyber-physical systems. It serves as a tool to assess the EMS and control functions from the cyber-physical perspective.

Future work includes assessing the impacts of measurement errors, topology errors, and communication delays on wide-area controls. Cybersecurity is also an important application that can be evaluated with the testbed.

### 7. Acknowledgments

This work was supported in part by the Engineering Research Centre Program of the National Science Foundation and the Department of Energy under NSF Award Number EEC-1041877 and the CURENT Industry Partnership Program. The authors also would like to acknowledge the efforts from Nan Duan, Denis Osipov and Dr. Mario Arrieta Paternina.

## 8. References

1. Tomsovic, K., Bakken, D.E., Venkatasubramanian, V., Bose, A.: 'Designing the next generation of real-time control, communication, and computations for large power systems' *Proc. IEEE*, 2005, **93**, (5), pp. 965–979.
2. Chow, J.H., Cheung, K.W.: 'A toolbox for power system dynamics and control engineering education and research' *IEEE Trans. Power Syst.*, 1992, **7**, (4), pp. 1559–1564.
3. Milano, F.: 'An open source power system analysis toolbox' *IEEE Trans. Power Syst.*, 2005, **20**, (3), pp. 1199–1206.
4. Cole, S., Belmans, R.: 'Matdyn, a new matlab-based toolbox for power system dynamic simulation' *IEEE Trans. Power Syst.*, 2011, **26**, (3), pp. 1129–1136.
5. Kelley, B.M., Top, P., Smith, S.G., Woodward, C.S., Min, L.: 'A federated simulation toolkit for electric power grid and communication network co-simulation', in '2015 Workshop on Modeling and Simulation of Cyber-Physical Energy Systems (MSCPES)' (2015), pp. 1–6
6. Vanfretti, L., Rabuzin, T., Baudette, M., Murad, M.: 'iTesla Power Systems Library (iPSL): A Modelica library for phasor time-domain simulations' *SoftwareX*, 2016.
7. Antonioli, D., Tippenhauer, N.O.: 'MiniCPS: A toolkit for security research on CPS Networks', in 'Proceedings of the First ACM Workshop on Cyber-Physical Systems-Security and/or Privacy' (2015), pp. 91–100
8. Hannon, C., Yan, J., Jin, D.: 'DSSnet: A Smart Grid Modeling Platform Combining Electrical Power Distribution System Simulation and Software Defined Networking Emulation' *Proc. 2016 Annu. ACM Conf. SIGSIM Princ. Adv. Discret. Simul. - SIGSIM-PADS '16*, 2016, pp. 131–142.
9. Adhikari, U., Morris, T.H., Pan, S.: 'A cyber-physical power system test bed for intrusion detection systems' *2014 IEEE PES Gen. Meet.*, 2014, pp. 1–5.
10. Ashok, A., Hahn, A., Govindarasu, M.: 'A Cyber-Physical Security testbed for Smart Grid : System Architecture and Studies' *Proc. Seventh Annu. Work. Cyber Secur. Inf. Intell. Res. ACM, 2011.*, 2011, p. 20.
11. Lin, H., Veda, S.S., Shukla, S.S., Mili, L., Thorp, J.: 'GECO: Global event-driven co-simulation framework for interconnected power system and communication network' *IEEE Trans. Smart Grid*, 2012, **3**, (3), pp. 1444–1456.
12. Dong, X., Lin, H., Tan, R., Iyer, R.K., Kalbarczyk, Z.: 'Software-Defined Networking for Smart Grid Resilience: Opportunities and Challenges' *Proc. 1st ACM Work. Cyber-Physical Syst. Secur.*, 2015, pp. 61–68.
13. Ghosh, U., Dong, X., Tan, R., Kalbarczyk, Z.: 'A Simulation Study on Smart Grid Resilience under Software-Defined Networking Controller Failures', in 'CPSS'16' (2016), pp. 52–58
14. Lin, H., Chen, C., Wang, J., Member, S., Qi, J., Jin, D.: 'Self-Healing Attack-Resilient PMU Network for Power System Operation' *IEEE Trans. Smart Grid*, 2016, **PP**, (99).
15. Kleissl, J., Agarwal, Y.: 'Cyber-physical energy systems: Focus on smart buildings' *Des. Autom. Conf. (DAC), 2010 47th ACM/IEEE*, 2010, pp. 749–754.
16. Sridhar, S., Hahn, A., Govindarasu, M.: 'Cyber-physical system security for the electric power grid' *Proc. IEEE*, 2012, **100**, (1), pp. 210–224.
17. Bergman, D.C., Jin, D., Nicol, D.M., Yardley, T.: 'The virtual power system testbed and inter-testbed integration' *Proc. 2nd Conf. Cyber Secur. Exp. test*, 2009, (August), p. 5.
18. Kundur, D., Feng, X., Liu, S., Zournos, T., Butler-Purry, K.L.: 'Towards a framework for cyber attack impact analysis of the electric smart grid' *SmartGridComm'2010*, 2010, pp. 244–249.
19. Shi, Q., Member, S., Cui, H., Member, S., Li, F.: 'A Hybrid Dynamic Demand Control Strategy for Power System Frequency Regulation' *Csee*, 2017, **3**, (2), pp. 176–185.
20. Zhao, J., Zhan, L., Liu, Y., Qi, H., Garcia, J.R., Ewing, P.D.: 'Measurement accuracy limitation analysis on synchrophasors', in '2015 IEEE Power & Energy Society General Meeting' (2015), pp. 1–5
21. OpenElectrical: 'Power Systems Analysis Software', http://www.openelectrical.org/wiki/index.php?title=Power_Systems_Analysis_Software, accessed January 2016
22. Hindmarsh, A.C., Brown, P.N., Grant, K.E., *et al.*: 'Sundials' *ACM Trans. Math. Softw.*, 2005, **31**, (3), pp. 363–396.
23. Lantz, B., Heller, B., McKeown, N.: 'A network in a laptop: rapid prototyping for software-defined networks', in 'Proceedings of the 9th ACM SIGCOMM Workshop on Hot Topics in Networks' (2010), p. 19
24. Abur, A., Rouhani, A.: 'Linear Phasor Estimator Assisted Dynamic State Estimation' *IEEE Trans. Smart Grid*, 2016, **3053**, (99), pp. 1–1.
25. Sun, K., Hur, K., Zhang, P.: 'A new unified scheme for controlled power system separation using synchronized phasor measurements' *IEEE Trans. Power Syst.*, 2011, **26**, (3), pp. 1544–1554.
26. Yuan, H., Li, F.: 'Hybrid voltage stability assessment (VSA) for N - 1 contingency' *Electr. Power Syst. Res.*, 2015, **122**, pp. 65–75.
27. 'CURENT: Center for Ultra-Wide-Area Resilient Electric Energy Transmission Networks', www.curent.utk.edu, accessed September 2017